\documentclass[amssymb,aps,twocolumn,floats,floatfix]{revtex4}
\usepackage{epsfig,color,graphicx,pstricks,bm}
\newcommand{\kk}{\ensuremath{{\bm{k}}}}
\newcommand{\ak}{\ensuremath{\alpha_{\bm{k}}}}
\newcommand{\bk}{\ensuremath{\beta_{\bm{k}}}}
\newcommand{\akdag}{\ensuremath{\alpha_{\bm{k}}^\dag}}
\newcommand{\bkdag}{\ensuremath{\beta_{\bm{k}}^\dag}}

\newcommand{\bone}{\ensuremath{\beta_1}}
\newcommand{\aonedag}{\ensuremath{\alpha_1^\dag}}

\newcommand{\btwo}{\ensuremath{\beta_2}}
\newcommand{\atwodag}{\ensuremath{\alpha_2^\dag}}

\newcommand{\athree}{\ensuremath{\alpha_3}}

\newcommand{\bthreedag}{\ensuremath{\beta_3^\dag}}
\newcommand{\afour}{\ensuremath{\alpha_4}}

\newcommand{\bfourdag}{\ensuremath{\beta_4^\dag}}
\newcommand{\wo}{\ensuremath{\widetilde{\omega}}}

\begin{document}

  \title{Spin transport in Heisenberg antiferromagnets}

  \author{M.\ Sentef\thanks{\email{michael.sentef@physik.uni-augsburg.de}},
    M.\ Kollar, and A.~P.\ Kampf}

  \affiliation{Theoretical Physics III,
    Center for Electronic Correlations and Magnetism,
    Institute of Physics,\\
    University of Augsburg, 86135 Augsburg, Germany}

  \date{December 8, 2006}

\begin{abstract}
We analyze spin transport in insulating antiferromagnets described by the XXZ Heisenberg model in
two and three dimensions. Spin currents can be generated by a magnetic-field gradient or, in systems
with spin-orbit coupling, perpendicular to a time-dependent electric field. The Kubo formula for the
longitudinal spin conductivity is derived analogously to the Kubo formula for the optical
conductivity of electronic systems. The spin conductivity is calculated within interacting spin-wave
theory. In the Ising regime, the XXZ magnet is a spin insulator. For the isotropic Heisenberg model,
the dimensionality of the system plays a crucial role: In $d=3$ the regular part of the spin
conductivity vanishes linearly in the zero frequency limit, whereas in $d=2$ it approaches a
finite zero frequency value.
\end{abstract}
\vskip0.15cm

\maketitle

\section{Introduction}
The challenge of spintronics research is to exploit the spin degree of freedom as an
additional tool in novel electronic devices \cite{Wolf,Awschalom,Slonczewski,Zutic,Koenig}. This
task demands to explore the basic physical principles underlying the generation and decay of
spin-polarized charge currents. In this context, metallic and semiconducting devices have been
examined both theoretically and experimentally. Simultaneously, the synthesis of
quasi-one-dimensional correlated insulators such as $\rm{Sr}_2\rm{CuO}_3$ \cite{Takigawa} and
$\rm{CuGeO}_3$ has renewed the general interest in the spin and thermal transport properties of
low-dimensional quantum spin systems. Also fundamental questions have been raised, e.g., how the
integrability of a system influences its spin \cite{Zotos,Alvarez,Fujimoto,Benz,Brenig} and thermal
\cite{Brenig,Saito,Kluemper,Jung} conductivities. In two-dimensional high-mobility electron systems
with Rashba spin-orbit coupling charge currents are necessarily accompanied by spin currents
\cite{Bychkov}. These spin currents flow perpendicular to the charge current direction and
therefore lead to an intrinsic spin Hall effect \cite{Murakami,Sinova}.

The spin conductivity of Heisenberg chains has previously been computed within linear-response
theory by adopting an analogy to the Kubo formula for charge transport \cite{Scalapino}. The
low-frequency behavior of the optical conductivity $\sigma(\omega) =
\sigma'(\omega)+\text{i}\sigma''(\omega)$ provides a transparent scheme to distinguish the
charge transport properties of ideal conductors, insulators, and non-ideal conductors
\cite{Scalapino,Souza}. Decomposing the real part of the longitudinal conductivity as
$\sigma'(\omega) = D\delta(\omega)+\sigma^{\rm{reg}}(\omega)$, a finite Drude weight or charge
stiffness \cite{Scalapino,Kohn} $D>0$ is the characteristic of ideal conductors. A similar
classification scheme can be carried over to spin transport and the spin conductivity in
order to distinguish between spin conductors, spin insulators or even spin superfluids
\cite{Coleman,Shastry}.

The optical conductivity is conveniently derived as the current response to a time-dependent
electromagnetic vector potential. Similarly, for the spin current response the concept of a
fictitious spin vector potential can be introduced, which is related to a twist in the direction of
the spin quantization axis \cite{Zhuo}. However, the physical realization of this perturbation and
its relation to externally applied magnetic or electric fields is not obvious. Indeed, spin currents
flow in response to a magnetic-field gradient. In this case, the analogy to the generation of
electric currents by a potential gradient is straightforwardly established for one-dimensional
systems by a Jordan-Wigner transformation \cite{Jordan,Fradkin,Eliezer}. This transformation maps
spin-$\frac{1}{2}$ operators to creation and annihilation operators of spinless fermions; a
magnetic-field gradient for the spins thereby translates into a potential gradient for the
fermions. 

In this paper, we do not appeal to the analogy to the charge current response. Rather, we derive the
Kubo formula for the spin conductivity of XXZ Heisenberg magnets directly (Sec.\ \ref{kuboformula}).
In particular, a magnetic-field gradient or, in the presence of spin-orbit coupling
\cite{Katsura}, a time-dependent electric field can be used to drive a spin current (Sec.\
\ref{electresp}). The longitudinal spin conductivity of the Heisenberg antiferromagnet in
two and three dimensions is computed using spin-wave theory  (Sec.\ \ref{spinwave}) for both
the non-interacting-magnon approximation (Sec.\ \ref{nonintmagnons}) and the ladder
approximation for repeated two-magnon scattering processes (Sec.\ \ref{ladderapprox}). In
particular, the low-frequency behavior of the spin conductivity will be analyzed in detail revealing
distinct differences between the spin transport properties of 2D and 3D antiferromagnets.

\section{Kubo formula}
\label{kuboformula}

Specifically, consider the antiferromagnetic XXZ Heisenberg model (HAFM)
\begin{equation}
\label{xxz}
	\mathcal{H} = J \sum_{\langle i,j\rangle} \left(\Delta S_i^z S_j^z + \frac{1}{2} (S_i^+
	S_j^- + S_i^-S_j^+)\right)
\end{equation}
with nearest-neighbor exchange coupling $J>0$, an anisotropy parameter $\Delta$, and local spin
operators $\bm{S}_i$ of length $S$. The sum over $\langle i,j\rangle$ extends over the
nearest-neighbor bonds on a $d$-dimensional hypercubic lattice with lattice constant $a=1$ and $N$
sites. An
external space- and time-dependent magnetic field $B^z(l,t)$ couples to the spin system via the
Zeeman energy. The time-dependent Hamiltonian thus reads
\begin{equation}
	\mathcal{H}(t) = \mathcal{H} - \sum_l S^z(l) h^z(l,t),
\end{equation}
where $\bm{h}(l,t)=h^z(l,t) \bm{e}_z=g\mu_B B^z(l,t) \bm{e}_z$.

The spin current density operator $j_{i\rightarrow j}$ for the magnetization transport from site
$i$ to site $j$ is defined via the continuity equation
\begin{equation}
\label{continuity}
	\partial_t S_i^z + \sum_j j_{i\rightarrow j} = 0.
\end{equation}
Here, $\sum_j j_{i\rightarrow j}$ is the lattice divergence of the local spin current density at
site $i$. The operator $j_x(l) = j_{l\rightarrow l+x}$ thus follows from Heisenberg's equation of
motion $\dot S_i^z=\text{i}[\mathcal{H},S_i^z]$ and Eq.\ (\ref{continuity}) as
\begin{equation}
\label{spincurrent}
	j_x(l) = \frac{J\text{i}}{2} \left(S_l^+ S_{l+x}^- - S_l^- S_{l+x}^+\right),
\end{equation}
where $l+x$ is the nearest-neighbor site of site $l$ in the positive $x$ direction.

In our subsequent calculation we assume long-range antiferromagnetic order oriented along the $z$
direction and thus restrict the analysis to a scalar spin current operator for the magnetization
transport. The definition of a proper spin current vector operator $\bm{I}_{i\rightarrow j}$ is
rather subtle, as discussed in detail in Ref.\ \cite{Schuetz}. In fact, if
$\langle\bm{S}_i\rangle$ and $\langle\bm{S}_j\rangle$ are not collinear, only the projection of
$\bm{I}_{i\rightarrow j}$ onto the plane spanned by the local order parameters
$\langle\bm{S}_i\rangle$ and $\langle\bm{S}_j\rangle$ may be interpreted as a physical transport
current. If $\langle\bm{S}_i\rangle$ and $\langle\bm{S}_j\rangle$ are collinear, however, the
magnetization transport is indeed correctly described by the scalar current density operator defined
in Eq.\ (\ref{spincurrent}). We note that the proper definition of a spin current operator is also a
controversial issue in the context of spin transport in semiconductors with spin-orbit coupling
\cite{Shi}, but these issues are of no concern for the purpose of our analysis.

The linear spin current response to an external magnetic-field gradient is
\begin{equation}
	\langle j_x(\bm{q},\omega)\rangle = \chi_{jS}(\bm{q},\omega) h^z(\bm{q},\omega),
\label{responseansatz}
\end{equation}
where the Fourier transforms in time and space are defined as
\begin{equation}
	A(\bm{q},\omega) = \sum_l \int_{-\infty}^{\infty} \text{d}t\;
	e^{\text{i}(\omega t-\bm{q}\cdot\bm{l})} A(l,t).
\end{equation}
In Eq.\ (\ref{responseansatz}) we have introduced the dynamic susceptibility
\begin{equation}
	\chi_{jS}(\bm{q},\omega) = \frac{\text{i}}{\hbar N} \int_{0}^{\infty} \text{d}t\;
	e^{\text{i}(\omega+i0^+)t}
	\langle[j_x(\bm{q},t),S^z(-\bm{q},0)]\rangle.
\end{equation}
Using the spatial Fourier transform of the continuity equation (\ref{continuity}),
\begin{equation}
\label{continuityfourier}
	\dot S^z(\bm{q},t) + \text{i}\bm{q}\cdot\bm{j}(\bm{q},t) = 0,
\end{equation}
and assuming a magnetic-field gradient of $B^z$ only along the $x$ direction, $\chi_{jS}$ is
transformed by partial integration as
\begin{eqnarray}
	\chi_{jS}(\bm{q},\omega) &=& \frac{\text{i}}{\hbar N}
	\frac{1}{\text{i}(\omega+i0^+)}\bigg[-\langle[j_x(\bm{q}),S^z(-\bm{q})]\rangle\\\nonumber
	&&+ \int_0^{\infty} \text{d}t\; e^{\text{i}(\omega+i0^+)t} \langle[j_x(\bm{q},t),\text{i}q_x
	j_x(-\bm{q},0)]\rangle\bigg].
\end{eqnarray}
The response formula therefore becomes
\begin{equation}
\label{grad}
	\langle j_x(\bm{q},\omega)\rangle = 
	-\frac{\langle-K_x\rangle - \Lambda_{xx}(\bm{q},\omega)}{\text{i}(\omega+\text{i}0^+)}
	\text{i}q_x h^z(\bm{q},\omega),
\end{equation}
where we have introduced the spin-flip part of the exchange interaction along the $x$ direction
\begin{equation}
	\langle K_x\rangle = \frac{1}{2\hbar N} \sum_{l} J \langle S_l^+ S_{l+x}^- +
	S_l^- S_{l+x}^+\rangle,
\end{equation}
and the longitudinal retarded current-current corre\-lation function,
\begin{equation}
\label{correlation}
	\Lambda_{xx}(\bm{q},\omega) = \frac{\text{i}}{\hbar N}\int_0^{\infty} \text{d}t\;
	e^{\text{i}(\omega+\text{i}0^+)t}
	\langle[j_x(\bm{q},t),j_x(-\bm{q},0)]\rangle.
\end{equation}

\begin{figure}[t!]
\centerline{\epsfig{file=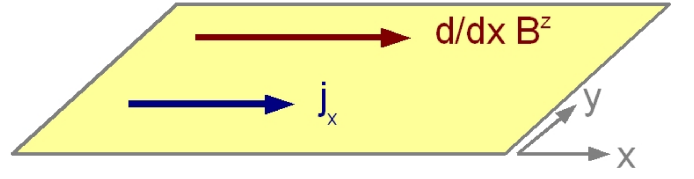,width=88mm,clip=true,silent=}}
\caption{(Color online) Setup for a spin current generated by a magnetic-field $B^z$ gradient along
the $x$ direction.}
\label{sentef1}
\end{figure}
For a closer analogy to charge transport, we consider the transport of
magnetization $m_l=g\mu_B S_l^z$ instead of (dimensionless) spin and define the
magnetization current operator $j_{m,x}=g\mu_B j_{x}$. The longitudinal spin conductivity
$\sigma_{xx}(\omega)$ is then defined as the linear magnetization current response to a
long-wavelength ($\bm{q}\rightarrow\bm{0}$), frequency dependent magnetic-field gradient (Fig.\
\ref{sentef1}). Putting $h^z=g\mu_B B^z$, the relation
\begin{equation}
	j_{m,x}(\bm{q},\omega)=\sigma_{xx}(\bm{q},\omega) \text{i}q_x B^z(\bm{q},\omega)
\end{equation}
thus yields the Kubo formula for the spin conductivity in the long-wavelength limit, 
\begin{equation}
\label{kubo}
	\sigma_{xx}(\omega) =
	-(g\mu_B)^2\frac{\langle-K_x\rangle -
	\Lambda_{xx}(\bm{q}=\bm{0},\omega)}{\text{i}(\omega+\text{i}0^+)}.
\end{equation}
This result indeed establishes the perfect analogy to the formula for the optical
conductivity of interacting lattice electrons \cite{Scalapino}. The real part of the spin
conductivity is given by
\begin{equation}
	\sigma'_{xx}(\omega) = D_S\delta(\omega) + \sigma_{xx}^{\text{reg}}(\omega)
\end{equation}
with the spin stiffness \cite{Coleman,Shastry} or ``spin Drude weight'' $D_S$,
\begin{eqnarray}
\label{drude}
	\frac{D_S}{\pi} = (g\mu_B)^2\big(\langle-K_x\rangle -
	\Lambda'_{xx}(\bm{q}=\bm{0},\omega\rightarrow0)\big),
\end{eqnarray}
and the regular part
\begin{equation}
	\sigma_{xx}^{\text{reg}}(\omega)=\frac{\Lambda''_{xx}(\bm{q}=\bm{0},\omega)}{\omega}.
\end{equation}
The spin conductivity also fulfills the ``f-sum rule'' \cite{Maldague}
\begin{equation}
	\frac{2}{\pi}\int_0^{\infty}\text{d}\omega\; \sigma'_{xx}(\omega)=\langle-K_x\rangle,
\label{sum}
\end{equation}
which can be derived by using the Kubo formula and the Kramers-Kronig relations for $\Lambda_{xx}$.
Note that the integral in Eq. (\ref{sum}) contains one half of the possible spin Drude weight peak
at zero frequency.

The structure of the spin conductivity formula emerges from the straightforward calculation for the
linear current response to an external magnetic-field gradient without introducing Peierls-like
phase factors with a fictitious spin vector potential \cite{Peierls,Alvarez}. However, in an
external electric field a moving magnetic dipole does acquire an Aharonov-Casher phase
\cite{Aharonov}. Katsura et al.\ \cite{Katsura} pointed out that Aharonov-Casher phase factors
and a corresponding ``vector potential'' can be introduced on the basis of the Dzyaloshinskii-Moriya
interaction \cite{Dzyaloshinskii,Moriya}, which may be induced in an external electric field due to
spin-orbit coupling. As we show in the next section, this alternative approach leads to the same
result for the spin conductivity, Eq.\ (\ref{kubo}).

\section{Response to an electric field}
\label{electresp}

Spin currents are also generated by applying a time-dependent electric field in the presence of
spin-orbit coupling. In low-symmetry crystals with localized spin moments spin-orbit coupling,
parametrized by a coupling constant $\lambda_{\text{so}}$, leads to the Dzyaloshinskii-Moriya
(DzM) antisymmetric exchange interaction
\cite{Dzyaloshinskii,Moriya}
\begin{eqnarray}
	\mathcal{H}_{\text{DzM}} = \sum_{\langle i,j\rangle} \bm{D}_{ij} \cdot (\bm{S}_i \times
	\bm{S}_j).
\end{eqnarray}
In high-symmetry crystals an inversion symmetry breaking external electric field $\bm{E}$ induces a
DzM vector $\bm{D}_{ij}\propto \bm{E} \times \bm{e}_{ij}$ \cite{Shiratori}, where $\bm{e}_{ij}$
denotes the unit vector connecting the neighboring sites $i$ and $j$. Specifically, for a field
$\bm{E}=(0,E_y,0)$, the DzM vector on the bonds in the $x$-$y$ plane takes the form
\begin{equation}
	\bm{D}_{i,i+x} = \alpha \bm{E} \times \bm{e}_x = -\alpha E_y \bm{e}_z,
	\quad \bm{D}_{i,i+y} = \bm{0},
\end{equation}
where $\alpha\propto \lambda_{\text{so}}$. With $(\bm{S}_i\times \bm{S}_j)^z = \frac{1}{2}
(S_i^+S_j^--S_i^-S_j^+)$, the DzM Hamiltonian can therefore be rewritten as
\begin{equation}
	\mathcal{H}_{\text{DzM}} = -\sum_{l} A_x(l,t) j_x(l),
\end{equation}
where $j_x(l)$ is the spin current density operator defined in Eq.\ (\ref{spincurrent}) and
\begin{equation}
	A_x(l,t) = \frac{2\alpha E_y(l,t)}{J}.
\end{equation}
The electric field therefore acts as a ``spin vector potential'' in formal analogy to the
electromagnetic vector potential in charge transport.

\begin{figure}[t!]
\centerline{\epsfig{file=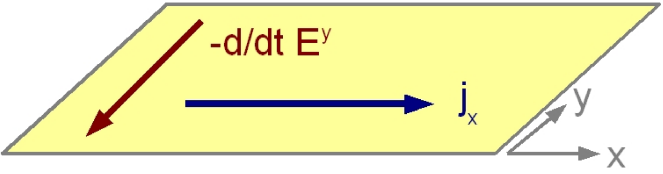,width=88mm,clip=true,silent=}}
\caption{(Color online) Setup for a transverse spin current driven by a time-dependent electric
field in the presence of spin-orbit coupling.}
\label{sentef2}
\end{figure}

If the DzM exchange interaction is added to the XXZ spin Hamiltonian, the Heisenberg equation of
motion yields an additional contribution to the spin current operator,
\begin{equation}
	j_{i\rightarrow j}^{\text{DzM}} =
	(D_{ji}-D_{ij}) \frac{1}{2} (S_i^+ S_j^- + S_i^- S_j^+),
\end{equation}
and the total magnetization current operator is given by
\begin{equation}
\label{extcurrent}
	j_{m,x}(l;A) = g\mu_B j_x(l;A=0) + (g\mu_B)^2 K_x(l) A_x(l).
\end{equation}
As for the charge current response to an electromagnetic vector potential \cite{Scalapino},
the linear spin current response to the spin vector potential in the long-wavelength limit becomes
\begin{equation}
	\langle j_{m,x}(\bm{0},\omega)\rangle = (g\mu_B)^2 \left(\langle K_x\rangle +
	\Lambda_{xx}(\bm{0},\omega)\right) A_x(\bm{0},\omega).
\end{equation}
The spin conductivity is therefore obtained by identifying the time derivative of the spin vector
potential, $\text{i}(\omega+\text{i}0^+)A_x(\bm{0},\omega)$, as the driving force for the spin
current. As a consequence, in the presence of spin-orbit coupling, a spin current can also be driven
perpendicular to a time-dependent electric field (see Fig.\ \ref{sentef2}). Physically, the origin
of this phenomenon is contained in the Maxwell equation
\begin{equation}
\label{maxwell}
	\text{rot}\;\bm{B} =
	\frac{4\pi}{c} \bm{j}_c + \frac{1}{c} \frac{\partial \bm{E}}{\partial t},
\end{equation}
in the absence of charge currents, $\bm{j}_c=\bm{0}$. Assuming a magnetic field in $z$ direction,
the $y$ component of Eq.\ (\ref{maxwell}) reduces to
\begin{equation}
	\partial_x B^z = \frac{-1}{c}\partial_t E_y = \frac{-J}{2\alpha c} \partial_t A_x.
\label{connection}
\end{equation}
Eq.\ (\ref{connection}) identifies the time-dependent electric field and the magnetic field gradient
as the same driving forces for the spin current.

\section{Spin-wave theory for the antiferromagnetic XXZ model}
\label{spinwave}

In our subsequent analysis of the dynamic spin current correlation function
$\Lambda_{xx}(\bm{0},\omega)$ we employ interacting spin-wave theory using the Dyson-Maleev (DM)
transformation \cite{Dyson,Maleev,Manousakis} for the antiferromagnetic Heisenberg model on
bipartite lattices. In the DM representation the spin operators are replaced by bosonic operators
according to
\begin{eqnarray}
	S_i^z &=& S - a_i^\dag a_i,\\
	S_i^+ &=& \sqrt{2S} \left(1-\frac{a_i^\dag a_i}{2S}\right) a_i,\quad
	S_i^- = \sqrt{2S} a_i^\dag \nonumber
\end{eqnarray}
for the up-spin sublattice $A$ and by
\begin{eqnarray}
	S_j^z &=& - S + b_j^\dag b_j ,\\
	S_j^+ &=& \sqrt{2S} b_j^\dag \left(1-\frac{b_j^\dag b_j}{2S}\right),\quad
	S_j^- = \sqrt{2S} b_j \nonumber
\end{eqnarray}
for the down-spin sublattice $B$. Due to the bosonic commutation relations for the $a$ and $b$
operators, the spin algebra is preserved.

The spatial Fourier transformation of the bosonic operators is conveniently written as \cite{Harris}
\begin{equation}
	a_i = \sqrt{\frac{2}{N}} \sum_\kk e^{-\text{i}\kk \cdot \bm{R}_i} a_{\kk},\;
	b_j = \sqrt{\frac{2}{N}} \sum_\kk e^{+\text{i}\kk \cdot \bm{R}_j} b_{\kk},
\end{equation}
where the momentum $\bm{k}$ is restricted to the magnetic Brillouin zone (MBZ). For the 2D
square lattice $\text{MBZ} =\{\kk: |k_x|+|k_y|\leq\pi\}$. Inserting the DM
representation into the XXZ Heisenberg model, the quadratic part of the resulting bosonic
Hamiltonian is diagonalized by the Bogoliubov transformation
\begin{equation}
	a_\kk = u_\kk \ak + v_\kk \bkdag,\quad 
	b_\kk = u_\kk \bk + v_\kk \akdag.
\end{equation}
The coefficients $u_\kk$ and $v_\kk$ are given by
\begin{equation}
	u_\kk = \sqrt{\frac{1+\varepsilon_\kk}{2\varepsilon_\kk}},\quad
	v_\kk = -\textrm{sgn}(\gamma_\kk)\sqrt{\frac{1-\varepsilon_\kk}{2\varepsilon_\kk}}.
\label{uv}
\end{equation}
The DM transformed Hamiltonian reads
\begin{equation}
\label{dysonhamiltonian}
	\mathcal{H}_{{\text{DM}}} = E_0 + H_0 + V_{{\text{DM}}}
\end{equation}
with the diagonalized quadratic part
\begin{equation}
\label{quadratic}
	\mathcal{H}_0 = \sum_\kk \hbar \Omega_\kk (\akdag\ak + \bkdag\bk).
\end{equation}
For a $d$-dimensional hypercubic lattice the spin-wave dispersion is
\begin{equation}
	\hbar\Omega_\kk = 2d\Delta JS\alpha(S)\varepsilon_\kk, \quad
	\varepsilon_\kk=\sqrt{1-\gamma_\kk^2/\Delta^2}
\end{equation}
with $\gamma_\kk = \frac{1}{d} \sum_{\alpha=1}^d \cos(k_\alpha)$. The magnon-vacuum energy is $E_0
= -N\Delta JS^2\alpha^2(S)d$. The Oguchi correction factor
\begin{equation}
\label{oguchi}
	\alpha(S) = 1+\frac{r}{2S},\quad
	r = 1-\frac{2}{N}\sum_\kk \varepsilon_\kk,
\end{equation}
arises from the normal-ordering of quartic terms \cite{Oguchi} in the interaction part
$V_{{\text{DM}}}$. In $d=2$,
\begin{equation}
	r = 1- {}_3F_2\left(-\frac{1}{2},\frac{1}{2},\frac{1}{2};1,1;\frac{1}{\Delta^2}\right),
\end{equation}
with the generalized hypergeometric function ${}_3F_2$ \cite{Abramowitz}. Specifically, for $d=2$,
$\Delta=1$, and $S=\frac{1}{2}$, the Oguchi correction factor is given by
\begin{equation}
	\alpha({\textstyle\frac{1}{2}}) = 2-
	\frac{8\pi}{\Gamma(\frac{1}{4})^4}-\frac{\Gamma(\frac{1}{4})^4}{8\pi}\approx 1.157947,
\end{equation}
where $\Gamma$ denotes the gamma function. Finally, the normal-ordered quartic interaction part of
the Dyson-Maleev Hamiltonian (\ref{dysonhamiltonian}) reads
\begin{eqnarray}
	&V_{\text{DM}}& = -\Delta J \frac{d}{N} \sum_{(1234)}
	\delta_{\bm{G}}(1+2-3-4)\\\nonumber
	&\times&\hspace{-4mm}\bigg[V^{(1)}\aonedag\atwodag\athree\afour+V^{(2)}
	\aonedag\btwo\athree\afour+V^{(3)}\aonedag\atwodag\bthreedag\afour\\\nonumber
	&&\hspace{-4mm}+V^{(4)}\aonedag\athree\bfourdag\btwo+V^{(5)}\bfourdag\athree\btwo\bone
	+V^{(6)}\bfourdag\bthreedag\atwodag\bone\\\nonumber
	&&\hspace{-4mm}+V^{(7)}\aonedag\atwodag\bthreedag\bfourdag+V^{(8)}
	\bone\btwo\athree\afour+V^{(9)}\bfourdag\bthreedag\btwo\bone\bigg].
\label{quartic}
\end{eqnarray}
The interaction vertices $V^{(i)}$, $i=1,\ldots,9$ depend on the wave vectors $\kk_1,\ldots,\kk_4$,
abbreviated by $1,\ldots,4$, and are explicitly given in Ref.\ \cite{Canali}. The Kronecker delta,
$\delta_{\bm{G}}(1+2-3-4)$, ensures momentum conservation to within a reciprocal lattice vector of
the MBZ.

We briefly comment on two difficulties with the Dyson-Maleev transformation. Firstly, two
constraints, $a_i^\dag a_i\leq2S$ and $b_j^\dag b_j\leq2S$, are required for the Fock
space of bosonic excitations in order to avoid unphysical spin excitations. The spin-wave analysis
without constraints can nevertheless be quantitatively justified noting that $\langle a^\dag_i
a_i\rangle\approx 0.197$ in $d=2$ at $T=0$ for the $S=\frac{1}{2}$ isotropic Heisenberg model,
calculated in linear spin-wave theory (LSW), where $V_{\text{DM}}$ is neglected. This result implies
for the sublattice magnetization $\langle m_{\text{A}}\rangle = \frac{1}{2}-\langle a^\dag_i
a_i\rangle\approx 0.303$ ($\Delta=1$), supporting a posteriori the validity of spin-wave theory even
for $S=\frac{1}{2}$.

Secondly, the Dyson-Maleev transformation is not Hermitian, since $S^+\neq (S^-)^\dag$
when the spin operators are expressed in terms of bosonic operators. As a consequence, the
quartic part of the transformed Hamiltonian is non-Hermitian, too. However, its non-Hermiticity does
not affect our calculations since we will take into account the only quantitatively relevant
Hermitian $V^{(4)}$-term (see Sec.\ \ref{ladderapprox}).

\section{Evaluation of the spin conductivity}
\label{computation}
\subsection{Basic propagators and correlation functions}

We proceed by defining the time-ordered
magnon propagators (see Refs. \cite{Canali,Harris} for more details),
\begin{eqnarray}
	G_{\alpha\alpha}(\kk,t)&\equiv&-\text{i}\langle0|\mathcal{T}
	\ak(t)\akdag(0)|0\rangle, \\
	G_{\beta\beta}(\kk,t)&\equiv&-\text{i}\langle0|\mathcal{T}
	\bkdag(t)\bk(0)|0\rangle.
\end{eqnarray}
The bare Fourier-transformed propagators in the absence of interactions are
\begin{equation}
	G_{\alpha\alpha}^{(0)}(\kk,\omega) = \frac{1}{\omega-\Omega_\kk+\text{i}0^+},\;
	G_{\beta\beta}^{(0)}(\kk,\omega) = \frac{-1}{\omega+\Omega_\kk-\text{i}0^+}.
\end{equation}
The regular part of the longitudinal spin conductivity
\begin{equation}
	\sigma'_{xx}(\omega) = (g\mu_B)^2 \frac{\Lambda''_{xx}(\bm{q=0},\omega)}{\omega} =
	-\frac{(g\mu_B)^2}{\omega}G''_j(\omega)
\label{regspincond}
\end{equation}
is determined by the imaginary part of the Fourier transformed time-ordered spin current correlation
function
\begin{eqnarray}
	G_j(t) \equiv -\frac{i}{\hbar N}\langle0|\mathcal{T}j_x(t)j_x(0)|0\rangle
\end{eqnarray}
in the ground state $|0\rangle$ with the spin current operator $j_x=\sum_l j_x(l)$ (see Eq.\
(\ref{spincurrent})). $j_x$ is transformed by the same steps which are used for the Dyson-Maleev
transformation of the HAFM Hamiltonian. The result is $j_x = j_0 + j_1$, where
\begin{eqnarray}
	j_0 &=& 2JS\alpha(S) \sum_\kk \sin(k_x)\\\nonumber
	&&\times\left[\frac{\gamma_\kk}{\Delta\varepsilon_\kk}
	(\akdag\ak+\bkdag\bk)-\frac{1}{\varepsilon_\kk} (\akdag\bkdag+\ak\bk)\right],
\label{j0}
\end{eqnarray}
and
\begin{eqnarray}
	&j_1& = J \frac{2}{N} \sum_{(1234)}
	\delta_{\bm{G}}(1+2-3-4)\sin(k_{1,x})\\\nonumber
	&\times&\hspace{-2mm}\bigg[j^{(1)}\aonedag\atwodag\athree\afour+j^{(2)}
	\aonedag\btwo\athree\afour+j^{(3)}\aonedag\atwodag\bthreedag\afour\\\nonumber
	&&\hspace{-2mm}+j^{(4)}\aonedag\athree\bfourdag\btwo+j^{(5)}\bfourdag\athree\btwo\bone
	+j^{(6)}\bfourdag\bthreedag\atwodag\bone\\\nonumber
	&&\hspace{-2mm}+j^{(7)}\aonedag\atwodag\bthreedag\bfourdag+j^{(8)}
	\bone\btwo\athree\afour+j^{(9)}\bfourdag\bthreedag\btwo\bone\bigg],
\label{j1}
\end{eqnarray}
where, for brevity, we have omitted here the explicit expressions of the spin current vertices
$j^{(i)}$; for completeness, they are listed in the Appendix. The quadratic part
$j_0$ is of order $S^1$, but it also contains Oguchi corrections of order $S^0$ arising from the
normal-ordering of quartic terms in $j_1$.

The calculation of $G_j(t)$ involves the following four correlation functions:

(i) A two-magnon correlation function
\begin{equation}
G_j^{00}(t) = -\text{i}\langle0|\mathcal{T}j_0(t)j_0(0)|0\rangle,
\end{equation}

(ii) a four-magnon correlation function
\begin{equation}
G_j^{11}(t) = -\text{i}\langle0|\mathcal{T}j_1(t)j_1(0)|0\rangle,
\end{equation}

(iii) and two cross-correlation functions
\begin{eqnarray}
G_j^{01}(t) = -\text{i}\langle0|\mathcal{T}j_0(t)j_1(0)|0\rangle,\\
G_j^{10}(t) = -\text{i}\langle0|\mathcal{T}j_1(t)j_0(0)|0\rangle.
\end{eqnarray}
The regular part of the spin conductivity (\ref{regspincond}) is then obtained with
\begin{equation}
	G_j(\omega) =
	\frac{1}{\hbar}\left[G_j^{00}(\omega)+G_j^{11}(\omega)+G_j^{01}(\omega)+G_j^{10}
	(\omega)\right].
\label{4correlations}
\end{equation}

Before we continue with the calculation of $G_j(\omega)$, we discuss the selection rules for the
matrix elements of the spin current operator, which provide insight into the relevant physical
processes for spin transport in antiferromagnets. The Lehmann representation
\begin{eqnarray}
\label{lehmann}
	&G_j(\omega)& = \frac{1}{\hbar N} \sum_n^{E_n\neq E_0}
	|\langle0|j_x|n\rangle|^2\\\nonumber
	&\times&\hspace{-2mm}\bigg[\frac{1}{\omega-(E_n-E_0)+\text{i}0^+}-\frac{1}{
	\omega+(E_n-E_0)-\text{ i} 0^+}\bigg],
\end{eqnarray}
shows that the spin current correlation function is determined by the matrix elements
$\langle0|j_x|n\rangle$, where $|n\rangle$ denotes the excited states with energy $E_n$.
The ground state and the relevant excited states must therefore have (i) the same total momentum and
(ii) the
same spin quantum numbers, but (iii) opposite parity in order to have nonvanishing matrix
elements $\langle0|j_x|n\rangle$. These conditions imply that only multi-magnon excitations with
vanishing total
momentum contribute to the spin conductivity; one-magnon excitations
are forbidden by both (i) and (ii). Condition (iii) is reflected in the $\sin(k_x)$ vertex in Eq.\
(\ref{j0}), which selects the $x$ direction for the spin current. Below we will focus on the
two-magnon contribution to the spin conductivity. Apart from the $\sin(k_x)$ vertex function,
our calculation proceeds analogously to the two-magnon analysis of Raman
scattering in antiferromagnets \cite{Canali}. In fact, the selection rules for spin transport
are the same as for the two-magnon Raman scattering intensity in $B_{1g}$ scattering geometry.

In the analysis of the spin current correlation function, Eq.\ (\ref{4correlations}), we
focus on $G_j^{00}$ because it provides the leading ($S^2$) order contribution and also the
dominant $S^1$ corrections for the spin conductivity. The product $j_0(t)j_0(0)$ contains 16 terms.
However, following the arguments of Canali and Girvin in Ref.\ \cite{Canali}, the contributions of
14 of these terms are negligibly small due to two arguments: Firstly, the 12 terms containing
prefactors $\gamma_\kk/\Delta\varepsilon_\kk$ are small because the absolute value of this factor is
small near the MBZ boundary, where the free magnon density of states has a van-Hove singularity.
Specifically, the four regions of the MBZ boundary where $|\sin(k_x)|\sim1$ provide the
quantitatively relevant contributions. Secondly, only two of the four remaining terms are nonzero if
magnon interactions are neglected.

The two remaining dominant terms lead to
\begin{equation}
	G_j(\omega) \approx G_j^+(\omega)+G_j^+(-\omega),
\label{twoterms}
\end{equation}
where
\begin{equation}
\label{gplus}
	G_j^+(\omega) = \frac{\left(2JS\alpha(S)\right)^2}{\hbar N}\sum_{\kk,\kk'}
	\frac{\sin(k_x)\sin(k'_x)}{\varepsilon_\kk\varepsilon_{\kk'}} \Pi_{\kk\kk'}(\omega).
\end{equation}
The two terms which are involved in Eq.\ \ref{twoterms} are thus calculated from the same
two-magnon Green function,
\begin{equation}
	\Pi_{\kk\kk'}(t) =
	-\text{i}\langle0|\mathcal{T}\ak(t)\bk(t)\alpha_{\kk'}^\dag(0)\beta_{\kk'}
	^\dag(0)|0\rangle.
\end{equation}
Its Fourier transform is expressed in terms of mag\-non propagators and a 3-point vertex function
\cite{Davies}
\begin{equation}
	\Pi_{\kk\kk'}(\omega)=\text{i}\int_{-\infty}^{\infty}\frac{\text{d}\omega'}{2\pi}
	G_{\alpha\alpha}(\kk,\omega+\omega')G_{\beta\beta}(\kk,\omega')
	\Gamma_{\kk\kk'}(\omega,\omega').
\end{equation}
The vertex function $\Gamma_{\kk\kk'}(\omega,\omega')$ satisfies the Bethe-Salpeter equation
\begin{eqnarray}
	\Gamma_{\kk\kk'}(\omega,\omega') &=& \delta_{\kk\kk'}-\frac{\text{i}}{\hbar}
	\frac{\Delta Jd}{N}\sum_{\kk_1}\int_{-\infty}^{\infty}\frac{\text{d}\omega_1}{2\pi}
	\nonumber\\
	&&\times\mathcal{V}_{\kk\kk_1\kk_1\kk}^{\alpha\beta}(\omega',\omega_1)
	G_{\alpha\alpha}(\kk_1,\omega+\omega_1)\nonumber\\
	&&\times G_{\beta\beta}(\kk_1,\omega_1)\Gamma_{\kk_1\kk'}(\omega, \omega_1).
\label{bethe0}
\end{eqnarray}
Here, the 4-point vertex $\mathcal{V}_{\kk\kk_1\kk_1\kk}^{\alpha\beta}(\omega',\omega_1)$ is the sum
of all the irreducible interaction parts \cite{Davies}.

In terms of magnon propagators and the vertex function, $G_j^+(\omega)$ is therefore given by
\begin{eqnarray}
	G_j^+(\omega) &=& 
	\frac{\text{i}\left(2JS\alpha(S)\right)^2}{\hbar N}
	\int_{-\infty}^{\infty}\frac{\text{d}\omega'}{2\pi}\sum_\kk
	\frac{\sin(k_x)}{\varepsilon_\kk}\nonumber\\
	&&\times
	G_{\alpha\alpha}(\kk,\omega+\omega')G_{\beta\beta}(\kk,\omega')\Gamma_{\kk}(\omega,\omega'),
\label{bethe}
\end{eqnarray}
where the reduced vertex function $\Gamma_{\kk}$ is defined by
\begin{equation}
	\Gamma_{\kk}(\omega,\omega') =
	\sum_{\kk'}\frac{\sin(k'_x)}{\varepsilon_{\kk'}}\Gamma_{\kk\kk'}(\omega,\omega').
\end{equation}
Eqs.\ (\ref{bethe0}) and (\ref{bethe}) define the integral equations which have to be solved in
the calculation of the correlation function $G_j^+(\omega)$. We will proceed in two steps: In a
first approximation, the interactions between magnons are omitted (Sec.\
\ref{nonintmagnons}). Subsequently, interactions are treated within a ladder
approximation for repeated magnon-magnon scattering processes (Sec.\ \ref{ladderapprox}).

\begin{figure}[t!]
\centerline{\epsfig{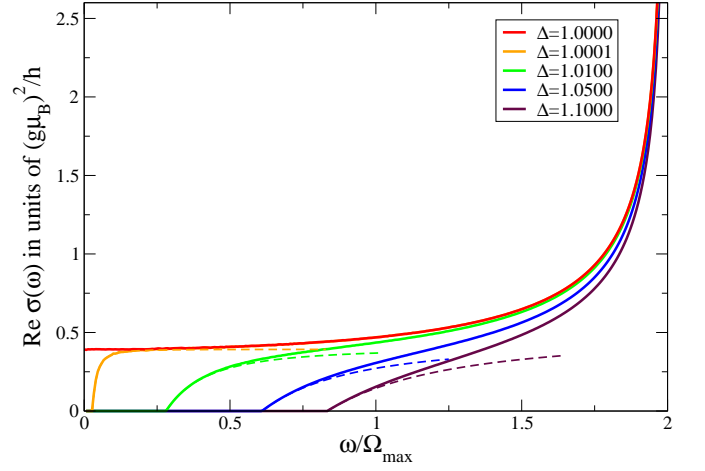}}
\caption{(Color online) Real part of the longitudinal spin conductivity for
non-interacting magnons in two dimensions for different values of the
anisotropy parameter $\Delta$. The dashed lines show the expansion around
$\omega_{\text{gap}}$, Eq.\ (\ref{gapexpansion1}).}
\label{sentef3}
\end{figure}
\subsection{Non-interacting magnons}
\label{nonintmagnons}

We start in a first step by neglecting magnon-magnon interactions, which amounts to the replacements
$G\rightarrow G^{(0)}$ and $\Gamma_{\kk}\rightarrow\sin(k_x)/\varepsilon_\kk$, i.e.\
$\mathcal{V}^{\alpha\beta}=0$. The required complex contour integral in Eq.\ (\ref{bethe0}) then
straightforwardly leads to the result
\begin{equation}
	G_j^+(\omega) =
	\frac{\left(JS\alpha(S)\right)^2}{\hbar N}\sum_\kk\frac{\sin^2(k_x)}{\varepsilon^2_\kk}
	\frac{1}{\omega-2\Omega_\kk+\text{i}0^+}.
\end{equation}
For convenience we introduce the dimensionless frequency $\wo = \omega/\Omega_{\text{max}}$,
where $\Omega_{\text{max}}=2d\Delta JS\alpha(S)/\hbar$ is the maximum one-magnon energy, and for
$m\in\{0,1,2\}$ we define the functions
\begin{equation}
	\ell^{(m)}(\wo)=
	\frac{2}{N} \sum_\kk\frac{\sin^2(k_x)}{\varepsilon_\kk^m}
	\frac{1}{\widetilde{\omega}-2\varepsilon_\kk+\text{i}0^+}.
\end{equation}
The regular part ($\wo>0$) of the spin conductivity within LSW is then given by
\begin{eqnarray}
\label{sigmanonint}
	\sigma'_{xx}(\omega) &=&
	-\frac{(g\mu_B)^2}{h}\frac{\pi}{(d\Delta)^2\wo}\;\text{Im}\;\ell^{(2)}(\wo)\\
	&=&\frac{(g\mu_B)^2}{h}\frac{\pi^2}{(d\Delta)^2\wo}\frac{2}{N}\sum_\kk\frac{\sin^2(k_x)}
	{\varepsilon_\kk^2}\delta(\widetilde{\omega}-2\varepsilon_\kk).\nonumber
\end{eqnarray}
Eq.\ (\ref{sigmanonint}) directly reflects the spin current selection rules explained above:
$\delta(\wo-2\varepsilon_\kk)$ accounts for two magnon excitations at
energy $\varepsilon_\kk$. Momentum conservation and spin conservation are fulfilled by a
combination of $\alpha$ and $\beta$ magnons, which carry opposite spin ($S^z=+1$ or $-1$) and
momentum ($\kk$ and $-\kk$). In a fully polarized Heisenberg ferromagnet, there is only one magnon
species and hence the spin conductivity vanishes \cite{Meier}.

The spin Drude weight (Eq.\ (\ref{drude})) $D_S=0$ for any $\Delta\geq1$ within LSW. Furthermore,
the regular part of the spin conductivity diverges at the maximum two-magnon energy ($\wo=2$) in
dimensions $d=2$ and $d=3$ due to the neglect of magnon interactions. The LSW result for
the spin conductivity in $d=2$ is shown in Fig.\ \ref{sentef3}. A special feature of the
regular part of $\sigma'_{xx}(\omega)$ is its finite zero frequency limit at the isotropic point
$\Delta=1$. An expansion near the magnon energy gap $\wo_{\text{gap}}\equiv2\sqrt{1-1/\Delta^2}$
yields the leading contribution as
\begin{eqnarray}
	\frac{\sigma'_{xx}(\omega)}{(g\mu_B)^2/h} &\simeq&
	\frac{\Delta^d d^{\frac{d}{2}-1}}{2^{\frac{3d}{2}}}
 	\frac{(\wo^2-\wo^2_{\text{gap}})^\frac{d}{2}}{\wo^2}
	\times
	\left\{\begin{array}{cl}
	\pi, & d=2 \\
	1, & d=3 \end{array}\right.\nonumber\\
\label{gapexpansion0}
\end{eqnarray}
for $\wo\geq\wo_{\text{gap}}$, with corrections of order $(\wo-\wo_{\text{gap}})^{\frac{d}{2}+1}$.
Specifically in $d=2$, the result of the expansion is
\begin{equation}
\label{gapexpansion1}
	\sigma'_{xx}(\omega) \simeq
	\sigma_{\text{LSW}}^* \Delta^2 
	\left(1-\frac{\omega^2_{\text{gap}}}{\omega^2}\right)
\end{equation}
for $\omega\geq\omega_{\text{gap}}\equiv\Omega_{\text{max}}\wo_{\text{gap}}$. The finite zero
frequency value of the spin conductivity for $\Delta=1$ is thus obtained as
\begin{equation}
	\sigma'_{xx}(\omega\rightarrow 0)=\sigma_{\text{LSW}}^*=\frac{(g\mu_B)^2}{h}\frac{\pi}{8}.
\end{equation}
The expansion (\ref{gapexpansion0}) for the spin conductivity within LSW in $d=2$ is also shown in
Fig.\ \ref{sentef3}. This expansion reveals that in $d=3$, as opposed to the two-dimensional case,
the spin conductivity for $\Delta=1$ vanishes linearly in the zero frequency limit within LSW
(see Fig.\ \ref{sentef5}). The unphysical divergence at the upper edge of the LSW spectrum will be
cured by including magnon-magnon interactions as discussed in the next subsection.

\begin{figure}[t!]
\centerline{\epsfig{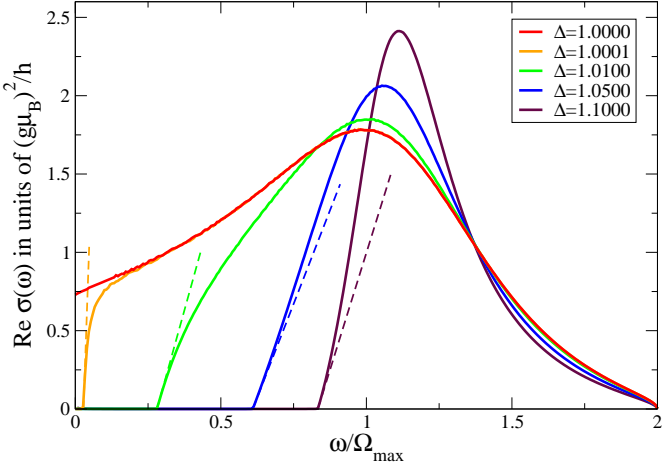}}
\caption{(Color online) Regular part of the longitudinal spin conductivity within the ladder
approximation for magnon-magnon interactions for $S=\frac{1}{2}$ in $d=2$. The dashed lines
represent the linear expansions around $\omega_{\text{gap}}$, Eq.\ (\ref{gapexpansion2}).}
\label{sentef4}
\end{figure}
\subsection{Ladder approximation: two-magnon scattering}
\label{ladderapprox}

For the calculation of $G_j^+(\omega)$ magnon-magnon interactions are taken into account to lowest
order \cite{Canali} by approximating the 4-point vertex $\mathcal{V}^{\alpha\beta}$ by its
first-order irreducible interaction part,
\begin{equation}
	\mathcal{V}_{\kk\kk_1\kk_1\kk}^{\alpha\beta}(\omega',\omega_1) = V_{\kk\kk_1\kk_1\kk}^{(4)},
\end{equation}
which is explicitly given by
\begin{equation}
	\frac{V_{\kk\kk_1\kk_1\kk}^{(4)}}{4} =
	\frac{\gamma_{\kk-\kk_1}}{2}\left(\frac{1}{\varepsilon_\kk\varepsilon_{\kk_1}}+1\right)-
	\frac{\gamma_\kk\gamma_{\kk_1}}{2\Delta^2\varepsilon_\kk\varepsilon_{\kk_1}},
\end{equation}
i.e., we neglect all the contributions to $\mathcal{V}^{\alpha\beta}$ where two or more of
the bare interactions $V^{(i)}$ are involved. Then the magnon propagators can again be replaced by
the bare expressions $G_{\alpha\alpha}^{(0)}$ and $G_{\beta\beta}^{(0)}$ in Eq.\ (\ref{bethe0})
since all the first order diagrams for the magnon self-energy vanish at $T=0$ \cite{Canali}.

The algebraic solution of the coupled integral equations (\ref{bethe0}) and (\ref{bethe}) is based
on the decoupling of the sums over $\kk$ and $\kk_1$ by means of the identities
\begin{eqnarray}
	\sum_\kk \sin(k_x)\gamma_\kk g_\kk &=& 0,\label{decoup1}\\
	\sum_\kk \sin(k_x)\gamma_{\kk-\kk_1} g_\kk &=& \frac{\sin(k_{1,x})}{d}\sum_\kk \sin^2(k_x)
	g_\kk,\label{decoup2}
\end{eqnarray}
which hold for any function $g_\kk$ which has the symmetry of the hypercubic lattice.

We proceed by repeatedly using Eqs.\ (\ref{decoup1}) and (\ref{decoup2}) to decouple Eqs.\
(\ref{bethe0}) and (\ref{bethe}) leading to a ladder approximation of the Bethe-Salpeter equation
for the vertex part $\Gamma_{\kk}(\omega,\omega')$. This has been demonstrated in Ref.\
\cite{Canali} for the vertex function $(\cos(k_x)-\cos(k_y))/2$ of the Raman $B_{1g}$ mode instead
of the spin current vertex $\sin(k_x)$. By virtue of the identities Eqs.\ (\ref{decoup1}) and
(\ref{decoup2}) the analogous analytical steps can be performed, and we obtain for the spin
conductivity
\begin{eqnarray}
\label{ladder}
	\sigma'_{xx}(\omega) &=& -\frac{(g\mu_B)^2}{h}\frac{\pi}{(d\Delta)^2\wo}\\
	&\times&\text{Im}\displaystyle
	\frac{\ell^{(2)}-\kappa(\ell^ {(1)}\ell^{(1)} - \ell^{(0)}\ell^{(2)})}
	{1+\kappa(\ell^{(0)}+\ell^{(2)})-\kappa^2(\ell^{(1)}\ell^{(1)}-\ell^{(0)}\ell^{(2)})},
	\nonumber
\end{eqnarray}
where $\ell^{(i)}\equiv\ell^{(i)}(\wo)$ and $\kappa^{-1}=2dS\alpha(S)$.

\begin{figure}[t!]
\centerline{\epsfig{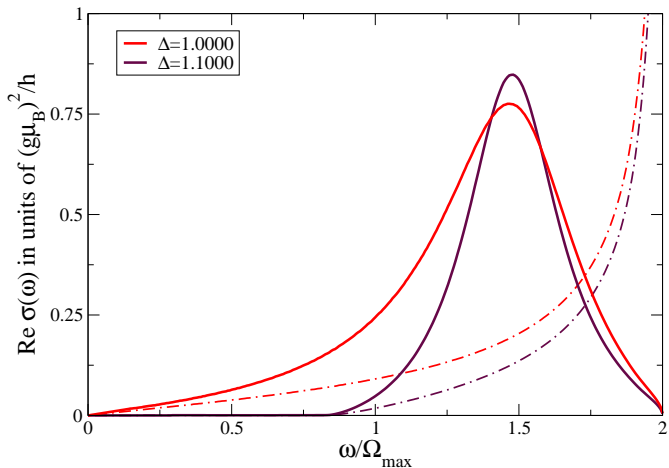}}
\caption{(Color online) Regular part of the longitudinal spin conductivity within the ladder
approximation for $S=\frac{1}{2}$ in $d=3$. The dashed lines show the LSW results.}
\label{sentef5}
\end{figure}

Figs.\ \ref{sentef4} and \ref{sentef5} show the $S=\frac{1}{2}$ spin conductivity within the ladder
approximation for $d=2$ and $d=3$, respectively. Scattering between magnons removes the divergence
of the LSW result at $\omega=2\Omega_{\text{max}}$. The frequencies for the maximal spin
conductivity increase with increasing anisotropy parameter $\Delta$.

In contrast to the spin conductivity of the 3D Heisenberg antiferromagnet, which vanishes at the
isotropic point $\Delta=1$ for $\omega\rightarrow0$, the most notable feature of the regular part of
the spin conductivity of the isotropic 2D Heisenberg antiferromagnet remains its finite value in the
zero frequency limit. For $\omega\geq\omega_{\text{gap}}$ we find the following leading-order linear
expansion in $d=2$,
\begin{equation}
\label{gapexpansion2}
	\sigma'_{xx}(\omega) \simeq
	\sigma_{\text{ladder}}^* A(\Delta)
	\left(1-\frac{\omega_{\text{gap}}}{\omega}\right),
\end{equation}
which is also included in Fig.\ \ref{sentef4}. $A(\Delta)$ is a numerical prefactor with $A(1)=1$,
and the zero frequency limit of the spin conductivity for $\Delta=1$ is given by
\begin{eqnarray}
	\sigma'_{xx}(\omega\rightarrow 0)&=&\sigma_{\text{ladder}}^*= Y_{\sigma}
	\frac{(g\mu_B)^2}{h}\frac{\pi}{8},\\
	Y_{\sigma} &\approx& 1.856851\quad\text{for}\; S=\frac{1}{2}.
\end{eqnarray}
The renormalization factor $Y_{\sigma}$ can be expressed in terms of the gamma function,
\begin{eqnarray}
	Y_{\sigma} &=&
	\frac{1-b_0\kappa+\frac{1}{4}b_0^2\kappa^2}{\left(1-\frac{1}{2}(b_0+b_2)\kappa-\frac{1}{4}
	\left(b_1^2-b_0b_2\right)\kappa^2\right)^2},\\\nonumber
	b_0 &=& \frac{32\pi}{\Gamma(\frac{1}{4})^4},\quad
	b_1 = 2-\frac{4}{\pi},\quad
	b_2 = \frac{\Gamma(\frac{1}{4})^4}{4\pi^3}-\frac{16\pi}{\Gamma(\frac{1}{4})^4}.
\end{eqnarray}
The inclusion of magnon-magnon interactions thus renormalizes $\sigma^*$ from its value for
non-interacting magnons. The fact that $\sigma'_{xx}(\omega)$ is finite in the limit
$\omega\rightarrow 0$, however, turns out to be a robust feature of the 2D Heisenberg
antiferromagnet.

\section{Summary and conclusions}
\label{conclusions}

In Heisenberg magnets, spin currents flow along a magnetic-field gradient or, in the
presence of spin-orbit coupling, perpendicular to a time-dependent electric field.
We presented an explicit derivation of the Kubo formula for the spin conductivity for the
antiferromagnetic XXZ Heisenberg model and showed that the magnetization transport arises
from two-magnon processes, which provide the dominant contribution to the spin conductivity.

In close analogy to the calculation of the $B_{1g}$ two-magnon Raman light scattering intensity, the
spin conductivity was evaluated using interacting spin-wave theory. The dimensionality of the model
is important for the low-frequency behavior of the spin conductivity, especially upon approaching
the isotropic point ($\Delta=1$) from the gapped Ising regime ($\Delta>1$). In $d=3$, the spin
conductivity vanishes for $\Delta=1$ in the dc limit. In $d=2$, however, the dc spin conductivity
remains \emph{finite} at the isotropic point.

Experimentally the spin conductivity can be determined by measurements of magnetization
currents. This issue was discussed by Meier and Loss in Ref.\ \cite{Meier}, where possible
experimental setups were proposed. In these setups the magnetization current is detected via the
electric field which it generates. Given the estimates in Ref.\ \cite{Meier} for the expected
voltages in moderate magnetic field gradients, measurements of the spin conductivity indeed appear
experimentally feasible.

\emph{Acknowledgment:} This work was supported by the Deutsche Forschungsgemeinschaft through
Sonderforschungsbereich 484.

\appendix*
\section{Spin current vertices}
The spin current vertices involved in Eq.\ (\ref{j1}) are given by
\begin{eqnarray*}
	j^{(1)} &=& u_1v_2u_3u_4+v_1u_2v_3v_4,\\
	j^{(2)} &=& u_1u_2u_3u_4+v_1v_2v_3v_4+v_1v_2u_3u_4+u_1u_2v_3v_4,\\
	j^{(3)} &=& 2(u_1v_2v_3u_4+v_1u_2u_3v_4),\\
	j^{(4)} &=& 2(u_1u_3v_4u_2+u_1v_3u_4u_2+\\
	&&v_1u_3v_4v_2+v_1v_3u_4v_2),\\
	j^{(5)} &=& 2(v_4u_3u_2v_1+u_4v_3v_2u_1),\\
	j^{(6)} &=& v_4v_3v_2v_1+u_4u_3u_2u_1+v_4v_3u_2u_1+u_4u_3v_2v_1,\\
	j^{(7)} &=& u_1v_2v_3v_4+v_1u_2u_3u_4,\\
	j^{(8)} &=& v_1u_2u_3u_4+u_1v_2v_3v_4,\\
	j^{(9)} &=& v_4v_3u_2v_1+u_4u_3v_2u_1.
\end{eqnarray*}
The coefficients $u_i=u_{\kk_i}$ and $v_i=v_{\kk_i}$ with $i=1,\ldots,4$ are defined by Eq.\
(\ref{uv}).


\begin{thebibliography}{99}
\bibitem{Wolf} S.~A.\ Wolf, D.~D.\ Awschalom, R.~A.\ Buhrman, J.~M.\ Daughton, S.\ von Moln\'{a}r, M.~L.\ Roukes, A.~Y.\ Chtchelkanova, and D.~M.\ Treger, Science {\bf 294}, 1488 (2001).

\bibitem{Awschalom} \textit{Semiconductor Spintronics and Quantum Computation}, edited by D.~D.\ Awschalom, D.\ Loss, and N.\ Samarth (Springer-Verlag, Berlin, 2002), and references therein.

\bibitem{Slonczewski} J.~C.\ Slonczewski, Phys.\ Rev.\ B {\bf 39}, 6995 (1989).

\bibitem{Zutic} The topic of spintronics was recently reviewed by I.\ \v{Z}uti\'{c}, J.\ Fabian, and S.\ {Das Sarma}, Rev.\ Mod.\ Phys.\ {\bf 76}, 323 (2004).

\bibitem{Koenig} J.\ K\"onig, M.~C.\ Bonsager, and A.~H.\ MacDonald, Phys.\ Rev.\ Lett.\ {\bf 87}, 187202 (2001).

\bibitem{Meier} F.\ Meier and D.\ Loss, Phys.\ Rev.\ Lett.\ {\bf 90}, 167204 (2003).

\bibitem{Takigawa} M.\ Takigawa, N.\ Motoyama, H.\ Eisaki, and S.\ Uchida, Phys.\ Rev.\ Lett.\ {\bf 76}, 4612 (1996).

\bibitem{Zotos} X.\ Zotos, Phys.\ Rev.\ Lett.\ {\bf 82}, 1764 (1999).

\bibitem{Alvarez} J.~V.\ Alvarez and C.\ Gros, Phys.\ Rev.\ Lett.\ {\bf 88}, 077203 (2002); Phys.\ Rev.\ B {\bf 66}, 094403 (2002).

\bibitem{Fujimoto} S.\ Fujimoto and N.\ Kawakami, Phys.\ Rev.\ Lett.\ {\bf 90}, 197202 (2003).

\bibitem{Benz} J.\ Benz, T.\ Fukui, A.\ Kl\"{u}mper, and C.\ Scheeren, J.\ Phys.\ Soc.\ Jpn.\ (Suppl.) {\bf 74}, 181 (2005).

\bibitem{Brenig} F.\ Heidrich-Meisner, A.\ Honecker, D.~C.\ Cabra, and W.\ Brenig, Phys.\ Rev.\ B {\bf 68}, 134436 (2003).

\bibitem{Saito} K.\ Saito, S.\ Takesue, and S.\ Miyashita, Phys.\ Rev.\ E {\bf 54}, 2404 (1996).

\bibitem{Kluemper} A.\ Kl\"{u}mper and K.\ Sakai, J.\ Phys.\ A {\bf 35}, 2173 (2002).

\bibitem{Jung} P.\ Jung, R.~W.\ Helmes, and A.\ Rosch, Phys.\ Rev.\ Lett.\ {\bf 96}, 067202 (2006).

\bibitem{Bychkov} Y.~A.\ Bychkov and E.~I.\ Rashba, J.\ Phys.\ C {\bf 17}, 6039 (1984).

\bibitem{Murakami} S.\ Murakami, N.\ Nagaosa, and S.\ Zhang, Science {\bf 301}, 1348 (2003).

\bibitem{Sinova} J.\ Sinova, D.\ Culcer, Q.\ Niu, N.~A.\ Sinitsyn, T.\ Jungwirth, and A.~H.\ MacDonald, Phys.\ Rev.\ Lett.\ {\bf 92}, 126603 (2004).

\bibitem{Scalapino} D.~J.\ Scalapino, S.~R.\ White, and S.\ Zhang, Phys.\ Rev.\ B {\bf 47}, 7995 (1993).

\bibitem{Souza} I.\ Souza, T.\ Wilkens, and R.~M.\ Martin, Phys.\ Rev.\ B {\bf 62}, 1666 (2000).

\bibitem{Kohn} W.\ Kohn, Phys.\ Rev.\ {\bf 133}, A171 (1964).

\bibitem{Coleman} P.\ Chandra, P.\ Coleman, and A.~I.\ Larkin, J.\ Phys.\ Cond.\ Mat.\ {\bf 2}, 7933 (1990).

\bibitem{Shastry} B.~S.\ Shastry and B.\ Sutherland, Phys.\ Rev.\ Lett.\ {\bf 65}, 243 (1990).

\bibitem{Zhuo} W.\ Zhuo, X.\ Wang, and Y.\ Wang, Phys.\ Rev.\ B {\bf 73}, 212413 (2006).

\bibitem{Jordan} P.\ Jordan and E.\ Wigner, Z.\ Phys.\ {\bf 47}, 631 (1928).

\bibitem{Fradkin} E.\ Fradkin, Phys.\ Rev.\ Lett.\ {\bf 63}, 322 (1989).

\bibitem{Eliezer} D.\ Eliezer and G.~W.\ Semenoff, Phys.\ Rev.\ B {\bf 286}, 118 (1992).

\bibitem{Katsura} H.\ Katsura, N.\ Nagaosa, and A.~V.\ Balatsky, Phys.\ Rev.\ Lett.\ {\bf 95}, 057205 (2005).

\bibitem{Schuetz} F.\ Sch\"{u}tz, P.\ Kopietz, and M.\ Kollar, Eur.\ Phys.~J.\ B {\bf 41}, 557 (2004).

\bibitem{Shi} J.\ Shi, P.\ Zhang, D.\ Xiao, and Q.\ Niu, Phys.\ Rev.\ Lett.\ {\bf 96}, 076604 (2006), and references therein.

\bibitem{Maldague} P.~F.\ Maldague, Phys.\ Rev.\ B {\bf 16}, 2437 (1977).

\bibitem{Peierls} R.~E.\ Peierls, Z.\ Phys.\ {\bf 80}, 763 (1933).

\bibitem{Aharonov} Y.\ Aharonov and A.\ Casher, Phys.\ Rev.\ Lett.\ {\bf 53}, 319 (1984).

\bibitem{Dzyaloshinskii} I.~E.\ Dzyaloshinskii, J.\ Phys.\ Chem.\ Solids {\bf 4}, 241 (1958).

\bibitem{Moriya} T.\ Moriya, Phys.\ Rev.\ Lett.\ {\bf 4}, 228 (1960); Phys.\ Rev.\ {\bf 120}, 91 (1960).

\bibitem{Shiratori} K.\ Shiratori and E.\ Kita, J.\ Phys.\ Soc.\ Jpn.\ {\bf 48}, 1443 (1980).

\bibitem{Canali} C.~M.\ Canali and S.~M.\ Girvin, Phys.\ Rev.\ B {\bf 45}, 7127 (1992).

\bibitem{Dyson} F.~J.\ Dyson, Phys.\ Rev.\ {\bf 102}, 1217 (1956); Phys.\ Rev.\ {\bf 102}, 1230 (1956).

\bibitem{Maleev} S.~V.\ Maleev, Zh.\ Eksp.\ Theor.\ Fiz.\ {\bf 30}, 1010 (1957) [Sov.\ Phys.\ JETP {\bf 64}, 654 (1958)].

\bibitem{Manousakis} E.\ Manousakis, Rev.\ Mod.\ Phys.\ {\bf 63}, 1 (1991).

\bibitem{Harris} A.~B.\ Harris, D.\ Kumar, B.~I.\ Halperin, and P.~C.\ Hohenberg, Phys.\ Rev.\ B {\bf 3}, 961 (1971).

\bibitem{Oguchi} T.\ Oguchi, Phys.\ Rev.\ {\bf 117}, 117 (1960).

\bibitem{Abramowitz} \textit{Handbook of Mathematical Functions}, edited by M.\ Abramowitz and I.~A.\ Stegun (Dover Publications, 1972).

\bibitem{Davies} R.~W.\ Davies, S.~R.\ Chinn, and H.~J.\ Zeiger, Phys.\ Rev.\ B {\bf 4}, 992 (1971).
\end{thebibliography}
\end{document}